\documentclass[preprint]{elsarticle} 
\usepackage{setspace}
\RequirePackage{amsmath}

\usepackage{graphicx}
\usepackage{subcaption}
 \usepackage{caption}
\usepackage{amsfonts}
\usepackage{booktabs}
\usepackage[numbers]{natbib}
\usepackage{notoccite}
\usepackage{eucal}
\usepackage{verbatim}
\usepackage{physics}

\usepackage{lineno}

\usepackage{color}
\usepackage{tikz}
\usetikzlibrary{shapes}

\usepackage{setspace}
\usepackage{bigints}
\usepackage{siunitx}
\usepackage{lipsum}
 \usepackage{xcolor}
 \usepackage{soul}
 
\usepackage{booktabs}
\usepackage{multirow}

\usepackage[english]{babel}
\usepackage{blindtext}
\usepackage{hyperref}
\hypersetup{
	unicode=true,
	pdftitle={Deep Reinforcement Learning in Action: Experimental Control of Vortex-Induced Vibrations},
	pdfnewwindow=true,
	colorlinks=true, 	
	pdfborder={0 0 0},
	linkcolor=blue,
	linktoc=all, 		
	citecolor=blue,
	urlcolor=blue,
	breaklinks=false,
}

\usepackage{nomencl}
\makenomenclature
\renewcommand{\nomgroup}[1]{%
  \ifthenelse{\equal{#1}{A}}{\item[\textbf{Dimensional variables}]}{%
  \ifthenelse{\equal{#1}{B}}{\item[\textbf{Dimensionless parameters}]}{%
  \ifthenelse{\equal{#1}{C}}{\item[\textbf{Reinforcement learning parameters}]}}}
}

\makeatletter
\let\save@mathaccent\mathaccent
\newcommand*\if@single[3]{%
  \setbox0\hbox{${\mathaccent"0362{#1}}^H$}%
  \setbox2\hbox{${\mathaccent"0362{\kern0pt#1}}^H$}%
  \ifdim\ht0=\ht2 #3\else #2\fi
  }
\newcommand*\rel@kern[1]{\kern#1\dimexpr\macc@kerna}
\newcommand*\widebar[1]{\@ifnextchar^{{\wide@bar{#1}{0}}}{\wide@bar{#1}{1}}}
\newcommand*\wide@bar[2]{\if@single{#1}{\wide@bar@{#1}{#2}{1}}{\wide@bar@{#1}{#2}{2}}}
\newcommand*\wide@bar@[3]{%
  \begingroup
  \def\mathaccent##1##2{%
    \let\mathaccent\save@mathaccent
    \if#32 \let\macc@nucleus\first@char \fi
    \setbox\z@\hbox{$\macc@style{\macc@nucleus}_{}$}%
    \setbox\tw@\hbox{$\macc@style{\macc@nucleus}{}_{}$}%
    \dimen@\wd\tw@
    \advance\dimen@-\wd\z@
    \divide\dimen@ 3
    \@tempdima\wd\tw@
    \advance\@tempdima-\scriptspace
    \divide\@tempdima 10
    \advance\dimen@-\@tempdima
    \ifdim\dimen@>\z@ \dimen@0pt\fi
    \rel@kern{0.6}\kern-\dimen@
    \if#31
      \overline{\rel@kern{-0.6}\kern\dimen@\macc@nucleus\rel@kern{0.4}\kern\dimen@}%
      \advance\dimen@0.4\dimexpr\macc@kerna
      \let\final@kern#2%
      \ifdim\dimen@<\z@ \let\final@kern1\fi
      \if\final@kern1 \kern-\dimen@\fi
    \else
      \overline{\rel@kern{-0.6}\kern\dimen@#1}%
    \fi
  }%
  \macc@depth\@ne
  \let\math@bgroup\@empty \let\math@egroup\macc@set@skewchar
  \mathsurround\z@ \frozen@everymath{\mathgroup\macc@group\relax}%
  \macc@set@skewchar\relax
  \let\mathaccentV\macc@nested@a
  \if#31
    \macc@nested@a\relax111{#1}%
  \else
    \def\gobble@till@marker##1\endmarker{}%
    \futurelet\first@char\gobble@till@marker#1\endmarker
    \ifcat\noexpand\first@char A\else
      \def\first@char{}%
    \fi
    \macc@nested@a\relax111{\first@char}%
  \fi
  \endgroup
}
\makeatother

\begin{document}
\begin{frontmatter}

\title{Deep Reinforcement Learning Discovers a Novel Control Algorithm for Mitigating Flow-Induced Vibrations in Underactuated Tandem Cylinders}

\author[address1]{Hussam Sababha\corref{mycorrespondingauthor}}
\cortext[mycorrespondingauthor]{Corresponding author}
\ead{haa385@nyu.edu}


\author[address1,address2]{Mohammed Daqaq}

\address[address1]{Division of Engineering, NYU Abu Dhabi, Abu Dhabi, UAE}
\address[address2]{Mechanical Engineering, Tandon School of Engineering, New York, USA}

\begin{abstract}
This study presents the first experimental implementation of deep reinforcement learning (DRL) for the active real-time suppression of flow-induced vibrations in simultaneously vibrating tandem cylinders using rotary actuation, considering fully actuated and underactuated configurations. In the fully actuated case, where both cylinders are independently controlled, the DRL agent discovers a high-frequency, phase-locked bang-bang control strategy that suppresses the vibrations of both cylinders by more than 95\%. Analysis of the training dynamics reveals a physically interpretable learning process in which the agent first identifies the optimal phase relationship between the actuators before refining the actuation frequency.

In the underactuated configuration, where only the upstream cylinder is actuated, equally weighted rewards produce ineffective control, suppressing vibrations only in the actuated cylinder. Introducing asymmetric reward weighting enables the DRL agent to discover a low-frequency lock-on strategy that achieves 70\% and 90\% vibration suppression in the upstream and downstream cylinders, respectively. For staggered arrangements with lateral offset, conventional training fails to converge, requiring a curriculum learning approach. The resulting two-stage curriculum identifies a statically biased bi-harmonic rotational control signal capable of suppressing vibrations in both cylinders. The success of the underactuated control strategy highlights its potential to reduce energy consumption and hardware complexity in multi-body flow control systems.
\end{abstract}


\end{frontmatter}




\nomenclature[A,01]{$M$}{Cylinder mass (kg)}
\nomenclature[A,02]{$D$}{Cylinder diameter (m)}
\nomenclature[A,03]{$K$}{Spring stiffness (N/m)}
\nomenclature[A,04]{$L$}{Immersed length (m)}
\nomenclature[A,05]{$Y$}{Transverse displacement (m)}
\nomenclature[A,06]{$A$}{Oscillation amplitude (m)}
\nomenclature[A,07]{$V$}{Free-stream flow velocity (m/s)}
\nomenclature[A,08]{$M_d$}{Displaced fluid mass (kg)}
\nomenclature[A,09]{$\rho$}{Fluid density (kg/m$^3$)}
\nomenclature[A,10]{$f_n$}{Natural frequency (Hz)}
\nomenclature[A,11]{$f_r$}{Rotational forcing frequency (Hz)} 
\nomenclature[A,12]{$\Omega$}{Cylinder rotational speed (rad/s)}
\nomenclature[A,13]{$\Omega_0$}{Reference rotational speed magnitude (rad/s)}
\nomenclature[A,14]{$T_0$}{Natural oscillation period (s)}
\nomenclature[A,15]{$\tau$}{Characteristic time scale (s)}
\nomenclature[A,16]{$T$}{Episode duration (s)}

\nomenclature[B,01]{$\alpha$}{Non-dimensional rotational speed}
\nomenclature[B,02]{$m$}{Mass ratio}
\nomenclature[B,03]{$\zeta$}{Damping ratio}
\nomenclature[B,04]{$U$}{Reduced velocity}
\nomenclature[B,05]{$SG$}{Skop–Griffin parameter}
\nomenclature[B,06]{$St$}{Strouhal number}

\nomenclature[C,01]{$s$}{State vector}
\nomenclature[C,02]{$a$}{Action }
\nomenclature[C,03]{$r$}{Reward function}
\nomenclature[C,04]{$\gamma$}{Discount factor}
\nomenclature[C,05]{$V(s)$}{State-value (advantage) function}
\nomenclature[C,06]{$Q(s,a)$}{State–action value function}
\nomenclature[C,07]{$\pi_\theta$}{Policy function parameterized by $\theta$}
\nomenclature[C,08]{$\theta$}{Neural network parameters}
\nomenclature[C,09]{$\epsilon$}{Clipping parameter for PPO}
\nomenclature[C,10]{$n$}{Number of past actions included in state history}

\section{Introduction}
\label{sec:introduction}
Flow-induced vibration (FIV) of closely spaced cylindrical rods is encountered in a broad range of engineering applications, including tube bundles in heat exchangers, marine risers, and energy-harvesting arrays~\cite{bernitsas2008vivace, weaver1988review, paidoussis1983review}. In such systems, the wake shed by each cylinder alters the flow field experienced by its neighbors, giving rise to complex fluid–structure interactions that differ fundamentally from those associated with isolated cylinders. These interactions produce a wide spectrum of instabilities, including vortex-induced vibration, wake-induced vibration, and wake galloping~\cite{papaioannou2008effect}, which can induce large-amplitude oscillations and jeopardize the structural integrity of the system. Accordingly, the suppression of such vibrations is of paramount importance and has been the subject of extensive investigation in the literature~\cite{xu2024review}.

To systematically address this challenge, prior work has often focused on a simplified configuration consisting of a pair of cylinders arranged in tandem along the streamwise direction. Despite its reduced complexity, this tandem arrangement preserves many of the essential features of wake coupling and has consequently emerged as a canonical benchmark for investigating multi-body fluid–structure interaction phenomena~\cite{sumner2010two}. Within this framework, a variety of control strategies have been explored. Among these, passive approaches that rely on geometric modifications of the rods have been proposed to reduce the hydrodynamic forces~\cite{kim2009flow, korkischko2010experimental, blumberg2012experimental, dongyang2018numerical}. While such methods are attractive due to their simplicity, they lack adaptability and are generally ineffective across varying flow conditions~\cite{xu2018flow}.

Active open- and closed-loop control strategies, that are based on injecting energy or momentum into the flow, have also been proposed to suppress FIV in single cylinders and tandem arrangements. In open-loop approaches, actuation is prescribed a priori without incorporating real-time information about the system dynamics~\cite{latrobe2024flow, eltaweel2014numerical, kozlov2011plasma}. In contrast, closed-loop strategies continuously adjust the actuation based on real-time measurements~\cite{rabiee2020simultaneous, wolfe2003feedback}. Among the two, closed-loop control is particularly promising, as it leverages instantaneous feedback to adapt to changing flow conditions, enabling robust performance across a wide range of operating regimes.

Despite its advantages, the application of active closed-loop control to tandem cylinder systems remains limited~\cite{xu2024review}. This limitation arises from two key challenges. First, wake coupling between the cylinders gives rise to complex nonlinear dynamics that are difficult to capture using reduced-order models or linearized stability analyses, rendering the design of classical model-based feedback controllers particularly challenging~\cite{soares2021modelling, mysa2016origin}. This difficulty is further exacerbated at high Reynolds numbers, where the wake becomes turbulent and three-dimensional~\cite{vojkovic2026parametric}. Second, the energy cost of active control increases with the number of actuated bodies. In multi-cylinder configurations, independently actuating each element quickly becomes impractical due to the associated hardware complexity and power requirements~\cite{chen2022review}.

Taken together, these challenges highlight the need for a closed-loop control framework with two key attribute. First, it should be model-free, thereby avoiding reliance on explicit governing equations or reduced-order models, which are difficult to construct for coupled multi-body wake dynamics. Second, it must be compatible with underactuated configurations, where only a subset of the bodies is directly actuated, while the remaining bodies are influenced indirectly through wake coupling.

Deep reinforcement learning (DRL) provides a promising pathway toward meeting these requirements~\cite{rabault2019artificial,Fan2020,Paris2021,font2025deep,Moslem2025}. As a model-free approach, DRL learns effective control policies through direct interaction with the flow environment, thereby bypassing the need for reduced-order models or explicit knowledge of the governing equations. Moreover, by leveraging neural networks as universal function approximators~\cite{hornik1989multilayer}, DRL can capture the nonlinear, time-delayed, and indirect coupling between an actuated body and its unactuated neighbors, which is difficult to achieve with classical control methods. 

Motivated by these advantages, DRL has recently been applied to the suppression of FIV in isolated cylinders. In one demonstration involving only computational simulations, Ren et al.~\cite{ren2024deep} applied DRL to discover a novel rotary control mechanism which suppresses vortex-induced vibration (VIV) of an elastically-mounted cylinder under lock-in conditions. In another demonstration, Sababha et al.~\cite{sababha2026deep} demonstrated a real-time experimental deployment of DRL for VIV suppression of an elastically-mounted cylinder at Re = 3000. Accounting for fundamental hardware constraints including actuator delay and limited sampling rates, their DRL-based control strategy achieved over 95\% vibration attenuation. 

More recently, a small number of studies have begun to explore DRL-based control for vibration suppression and drag reduction in tandem cylinder configurations~\cite{he2025deep, he2025jet, ren2024active, xie2023applying}. However, these studies share common limitations. First, the control task in each of these studies reduces to a single body problem as only the downstream cylinder was allowed to vibrate while the upstream cylinder is rigidly fixed. Second, these investigations are numerical in nature, and they are restricted to two-dimensional simulations at low Reynolds numbers.

To address these gaps, the present study conducts an experimental investigation of DRL-based rotary active control for vibration suppression in a tandem cylinder system in which both cylinders are elastically mounted and free to oscillate. This fully coupled configuration requires the DRL agent to simultaneously regulate the mutually interacting dynamics of two bodies, posing a substantially more challenging problem than the control of a single vibrating cylinder in the wake of a fixed one.

The study is structured around two configurations of increasing complexity and decreasing actuation authority. First, a fully actuated configuration, where both cylinders are independently controlled, is used to establish the feasibility of model-free DRL for multi-body FIV suppression and to elucidate the control strategies that emerge when the agent has direct control over both bodies. This is followed by an underactuated configuration, in which rotary actuation is applied to only one cylinder, requiring the DRL agent to suppress the vibrations of both cylinders using a single actuator. The success of the underactuated approach is particularly important to demonstrate that wake coupling provides a physical pathway through which the influence of a single actuator can propagate to unactuated ones. This paradigm can substantially reduce actuator count, energy consumption, and hardware complexity, making it especially well-suited for large-scale systems such as riser bundles, energy-harvesting arrays, and autonomous underwater vehicles~\cite{hong2018vortex, awadallah2025improving, li2026vortex}.

Throughout this study, our primary objective is to demonstrate that, with appropriate modifications to the training algorithms, DRL can experimentally uncover effective and nontrivial control strategies that are difficult to derive through conventional, physics-based modeling of such a complex system. A detailed investigation of how vortex shedding patterns organize and influence the resulting vibrations falls outside the scope of this already comprehensive work. Nevertheless, whenever feasible, we interpret the learned control strategies in the context of established fluid mechanics literature on rotary actuation and tandem cylinder configurations.

The remainder of this paper is organized as follows. Section~\ref{sec:problem} describes the problem statement. Section~\ref{sec:method} details the experimental setup, including the fluid--structure apparatus and sensing and actuation systems. Section~\ref{sec:validation} presents validation against published results of the uncontrolled tandem cylinder. Section~\ref{sec:DRL} presents the deep reinforcement learning framework, including the state and action representations, reward formulation, and training procedure. Section~\ref{sec:results} presents the results and discussion findings. Finally, Section~\ref{sec:conclusions} summarizes the key outcomes and outlines directions for future research.

\section{Problem Statement}
\label{sec:problem}
\begin{figure}[tb]
    \centering
\includegraphics[width=0.75\linewidth]{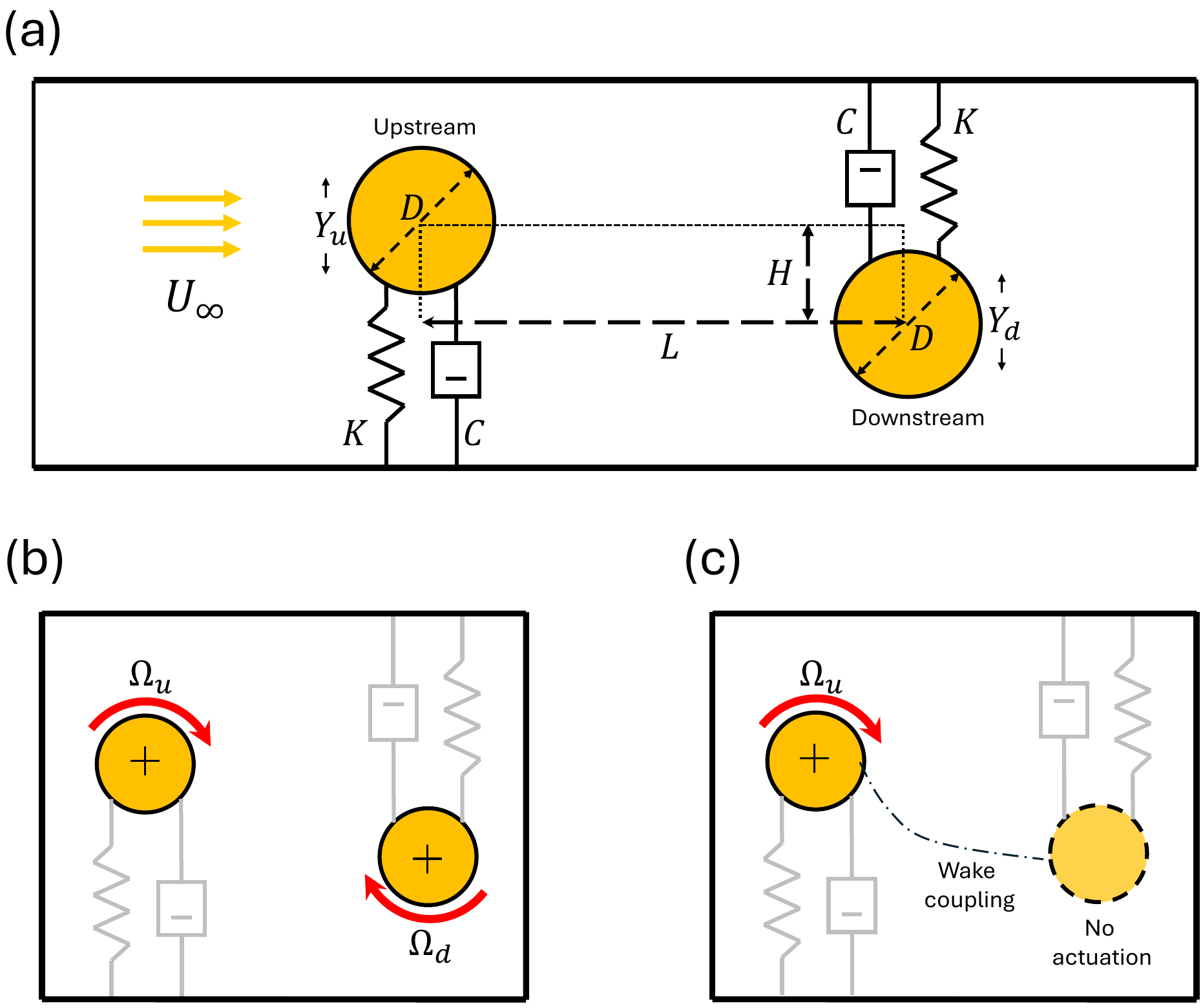}
\caption{(a) A schematic representation of two identical circular
cylinders of diameter, $D$, arranged in tandem with center-to-center
spacing, $L$, and lateral offset, $H$, subjected to a uniform flow of
velocity $U_\infty$. Each cylinder is free to oscillate in the
transverse, $Y$, direction and can be actively rotated about its axis
with angular velocity $\Omega$. Subscripts $"u"$ and $"d"$ denote the
upstream and downstream cylinders, respectively. (b) Fully actuated configuration in which both the upstream and downstream cylinders are independently actuated. (c) Underactuated system in which only the upstream is actuated while the downstream cylinder remains passive.}  
\label{fig:schematic}
\end{figure}

We consider the flow configuration illustrated in Fig.~\ref{fig:schematic}(a), consisting of two identical circular cylinders of diameter, $D$, elastically mounted in a uniform incoming flow of density, $\rho$, and free-stream velocity, $U_\infty$. The cylinders are aligned along the streamwise direction with a fixed center-to-center spacing, $L$, and offset distance, $H$. Each cylinder is constrained to oscillate in the transverse (cross-flow) direction, $Y$, with its internal dynamics governed by unknown structural mass, $M$, stiffness, $K$, and damping coefficient, $C$. Note that, throughout this work, the subscripts $``u"$ and $``d"$ denote the upstream and downstream cylinders, respectively.

In the absence of control, the upstream cylinder sheds a periodic vortex street that impinges on the downstream body, generating wake patterns, amplified force fluctuations, and large-amplitude oscillations~\cite{sumner2010two}. To suppress these oscillations, each cylinder may be prescribed an angular velocity, $\Omega$, about its axis. The resulting surface rotation injects tangential momentum into the boundary layer, modifying the vortex formation, and thereby altering the forces acting on both bodies.

The objective of the present work is to design a model-free closed-loop control strategy that suppresses the FIV of the two cylinders utilizing rotary actuation. The controller must operate in real time, mapping instantaneous observations of the state of the system comprising the displacement histories of both cylinders to continuous motor commands that suppress, or at the least minimize, the vibration amplitudes of both bodies. No model of the flow physics or structural dynamics will be utilized. This problem is investigated through two configurations of increasing complexity, see Figs.~\ref{fig:schematic}(b, c):
\begin{enumerate}
    \item \textbf{Fully actuated}: Both cylinders are independently controlled via rotary actuation; i.e., $\Omega_u \neq 0$, and $\Omega_d \neq 0$. This configuration serves as a baseline, establishing the maximum achievable suppression when the controller has direct authority over both bodies.
    \item \textbf{Underactuated}: The upstream cylinder is actuated, while the downstream cylinder is not actuated; i.e., $\Omega_u \neq 0$, and $\Omega_d = 0$. The downstream cylinder is influenced by modifications of the flow resulting from the motion of the upstream cylinder. In this case, the controller must exploit wake coupling to suppress the FIV of a body over which it has no direct authority. This 
    configuration is examined for both the inline ($H/D = 0$) and staggered ($H/D = 1$) arrangements.
\end{enumerate}

\section{Experimental Setup}
\label{sec:method}
\begin{figure}[tb]
    \centering
    \includegraphics[width=0.75\linewidth]{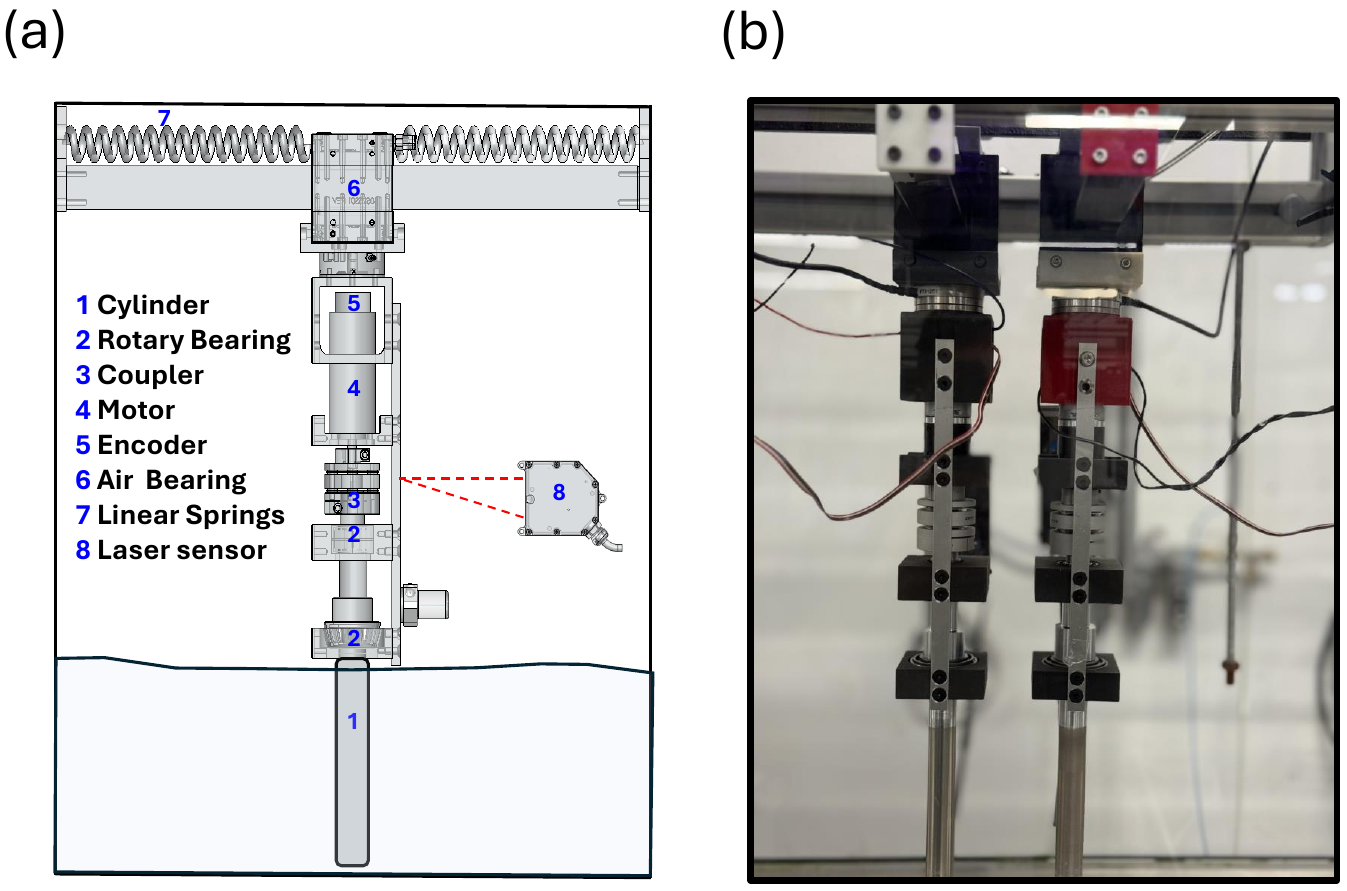}
    \caption{Experimental setup. (a) Schematic of a single cylinder assembly. (b) Full tandem installation mounted in the water channel.}
    \label{fig:experiment}
\end{figure}

The experimental system is created to mimic the conditions stated in the problem statement, where both cylinder assemblies are designed to be nominally identical. Figure~\ref{fig:experiment}(a) shows a schematic of a single cylinder assembly, while Fig.~\ref{fig:experiment}(b) presents the full tandem system.

Each test cylinder, denoted by \textcolor{blue}{1} in Fig.~\ref{fig:experiment}(a), is machined from Aluminum and has a diameter $D = 17.5 \pm 0.01~\si{mm}$ and an immersed length of $L = \SI{160}{mm}$, yielding an aspect ratio $L/D \approx 9.14$. Each cylinder is free on one end (the immersed end) and mounted on the other end onto a custom housing using two precision rotary bearings ``\textcolor{blue}{2}''. The rotary bearings, are rigidly mounted to a custom housing using alignment rods, which carry the weight and fix the cylinder along a defined rotational axis. The mounted end of the test cylinder is connected to a motor ``\textcolor{blue}{4}'' using a flexible coupler ``\textcolor{blue}{3}'', which mitigates any residual misalignment between the motor shaft and the rotating cylinder. The whole system including the test cylinder, the rotary motor, the rotary bearings and the housing are mounted on a linear track using a linear air bearing ``\textcolor{blue}{6}''. The air-bearing ensures low structural damping while constraining the motion solely in the cross-flow direction.  The air-bearing is connected on either side to two linear springs ``\textcolor{blue}{7}'' that provide the restoring force necessary to incite VIV.

The total mass of the oscillating system, encompassing the cylinder, motor assembly, and movable components of the air-bearing carriage, is identical for both cylinders, $M = \SI{1095}{g}$. The displaced fluid mass is $M_d = \rho \frac{\pi}{4} D^2 L_s = \SI{33.07}{g}$, yielding a mass ratio of $m^* = M/M_d = 33.04$. Free-decay tests conducted in quiescent water reveal a slight asymmetry between the two systems. The upstream cylinder has a natural frequency $f_{n,u} = \SI{1.35}{Hz}$ and a structural damping ratio $\zeta_u = 0.001$, while the downstream cylinder has $f_{n,d} = \SI{1.31}{Hz}$ and $\zeta_d = 0.0025$\footnote{It is worth noting that these experimentally identified parameter values are not utilized by the DRL algorithm in any capacity. They are reported solely for completeness and to enable reproducibility by other researchers.}. This minor discrepancy arises from manufacturing tolerances in the assemblies. 

Rotary actuation in both cylinders is provided by brushed DC motors with a rated torque of $0.24~\si{kg\cdot cm}$, mounted coaxially with each cylinder. The motors are driven by a Pulse Width Modulation (PWM) based controller that allows independent and continuous regulation of the angular velocities of the cylinders. The transverse displacement of each cylinder is measured by non-contact laser displacement sensors (micro-epsilon) denoted by ``\textcolor{blue}{8}'' on Fig.~\ref{fig:experiment}(a). The rotational speed of each motor is monitored by incremental rotary encoders mounted on the motor shafts (EP40) denoted by ``\textcolor{blue}{5}''. Motor actuation, data acquisition, and signal processing are fully integrated within a Python-based control framework. This unified environment enables synchronous execution of control commands and acquisition of sensor measurements at each control step, and serves as the direct interface between the physical system and the reinforcement learning agent.

\section{Response in The Absence of Control}
\label{sec:validation}
\begin{figure}[tb]
    \centering
    \includegraphics[width=0.80\linewidth]{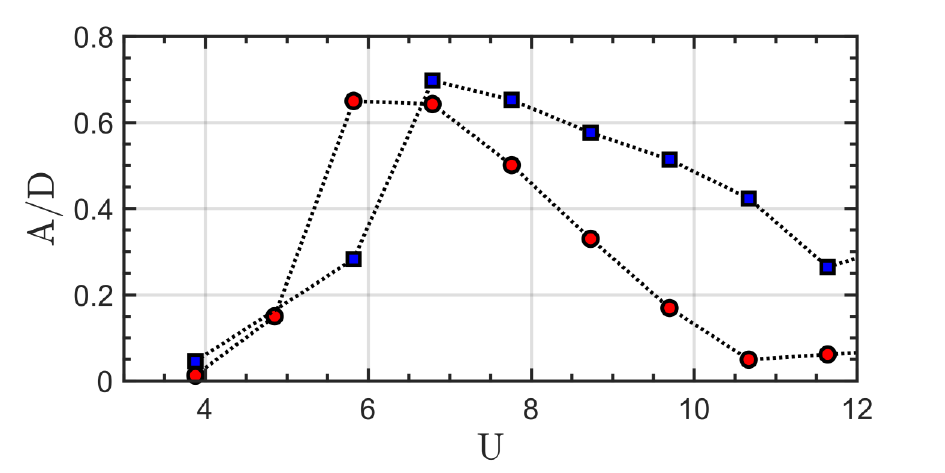}
    \caption{Dimensionless steady-state vibration amplitude, $A/D$, of the upstream (circles) and the downstream (squares) cylinders as a function of reduced velocity, $U$.}
    \label{fig:validation}
\end{figure}

Prior to designing the DRL controller, we study the steady-state FIV response of the tandem cylinder system in the absence of prescribed rotary control. For this study, the center-to-center distance between the two cylinders is set to $L = 5.5 D$ and the offset distance to $H = 0$. The reduced flow velocity, $U= U_\infty/(f_n D)$, is quasi-statically increased from $4$ to $12$ and the steady-state amplitude of oscillations of both cylinders is measured and recorded. Figure~\ref{fig:validation} presents variation of the normalized steady-state amplitude $A/D$ of both cylinders as a function of $U$.

The upstream cylinder response (red circles) exhibits the characteristic features of VIV observed in isolated cylinders. The amplitude remains negligible at low reduced velocities, increases sharply as it approaches the lock-in regime near $U = 4$, and reaches a peak of $A/D \approx 0.65$ near $U = 5.5$, before gradually decaying at higher reduced velocities. This behavior is qualitatively consistent with the classical VIV response reported for single cylinders~\cite{khalak1997investigation}. It is evident that, at the present spacing ratio, $L$, the upstream cylinder behaves largely as an independent oscillator whose dynamics are not influenced by the downstream cylinder, which is consistent with the numerical results of Prasanth and Mittal~\cite{prasanth2009flow}.

The response of the downstream cylinder (blue squares) follows a different behavior. In particular, the peak in the steady-state response amplitude occurs at a slightly higher reduced velocity of nearly $U = 6.5$ and exceeds that of the upstream cylinder, reaching a value of $A/D \approx 0.70$. This crossover, where the downstream cylinder vibrates with a larger amplitude compared to the upstream cylinder above a threshold reduced velocity, is a well-documented characteristic of FIV in tandem cylinders. It has been attributed to wake-induced vibrations (WIV) that involve gap flow effects and the impingement of upstream vortices on the downstream body~\cite{papaioannou2008effect}. The elevated response of the downstream cylinder persists over a wide range of reduced velocities ($5.5 \leq U \leq 12$), remaining substantially above that of the upstream cylinder. This is a signature of WIV, which is known to extend beyond the classical lock-in regime~\cite{assi2010wake}. 

At $U = 7$, both cylinders exhibit pronounced vibration amplitudes ($A/D \approx 0.60$ for the upstream cylinder and $A/D \approx 0.70$ for the downstream cylinder), making this operating condition particularly well suited for assessing the effectiveness of the DRL-based control strategy. Accordingly, all subsequent control experiments are conducted at this reduced velocity.

\section{\label{sec:DRL}Deep Reinforcement Learning}

\begin{figure}[tb]
    \centering
    \includegraphics[width=1.0\linewidth]
    {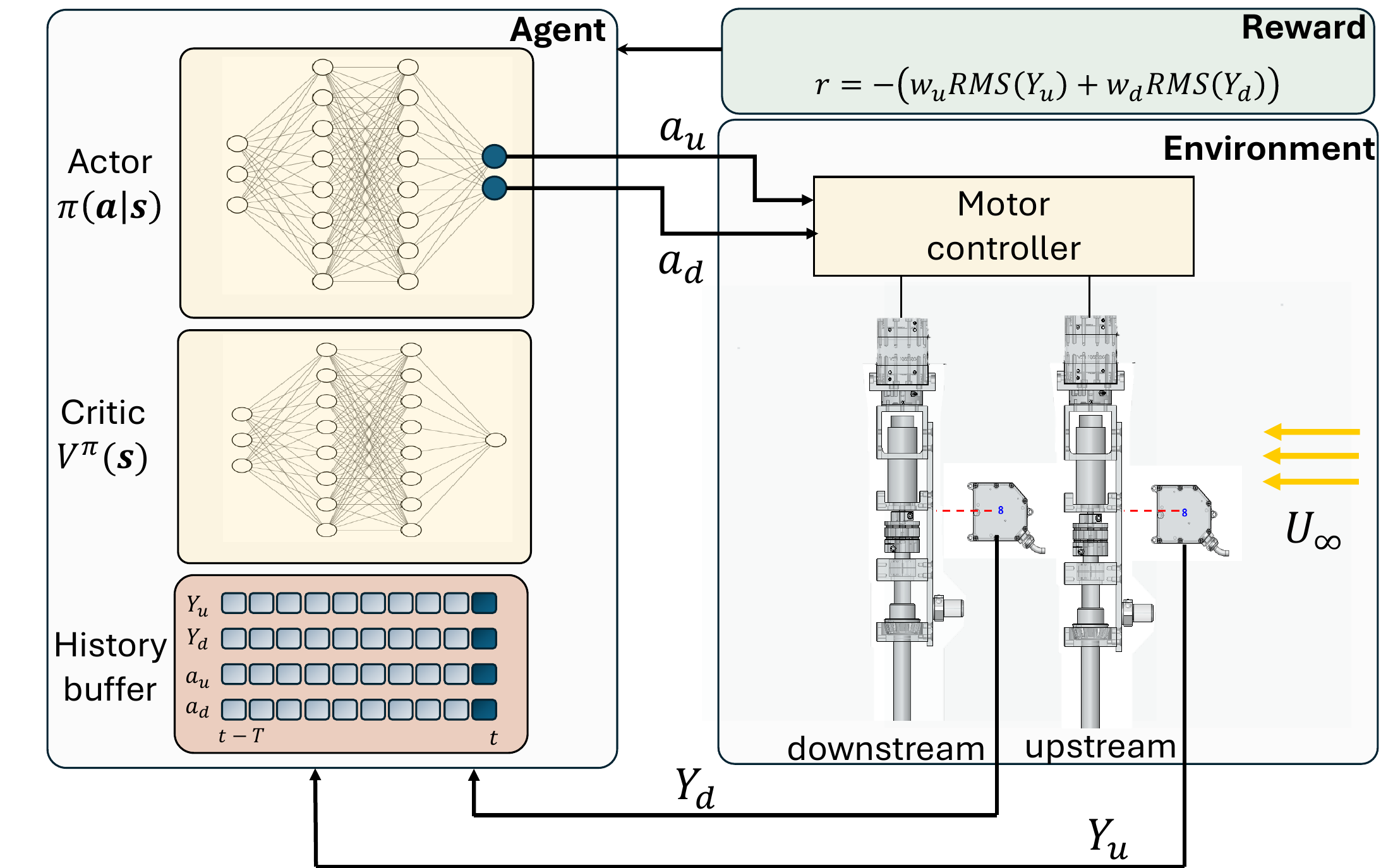}
    \caption{Schematic of the deep reinforcement learning (DRL) 
    framework. The agent observes the system state, $Y_i$, selects a 
    control action, $a_i$, and receives a scalar reward, $r$, from the 
    physical environment at each time step.}
    \label{fig:DRL}
\end{figure}

Unlike classical active control methods in fluid mechanics, which typically rely on reduced-order models or prior knowledge of the governing equations, deep reinforcement learning (DRL) learns control policies through direct interaction with the system, requiring no explicit model of the flow physics or structural dynamics. In the present work, a DRL agent interacts iteratively with the physical experimental setup and adapts its control strategy to minimize the vibrations of both cylinders simultaneously. The control framework is illustrated in Fig.~\ref{fig:DRL}.

The agent is trained using Proximal Policy Optimization (PPO)~\cite{ppo}, an actor--critic algorithm in which two neural networks are jointly trained. The \emph{actor} (policy network) maps observed states, $\textbf{s}=(Y_u, Y_d)$, to control actions, $\textbf{a}=(a_u, a_d)$, and the \emph{critic} (value network) estimates the expected cumulative reward from a 
given state. Both networks share an identical architecture consisting of two fully connected hidden layers with 256 units each and hyperbolic tangent activation functions, with weights initialized using orthogonal initialization to promote stable learning dynamics. The actor outputs the parameters of a 
Gaussian distribution $\pi_\theta(\mathbf{a}|\mathbf{s}) = \mathcal{N}(\mu_\theta(\mathbf{s}), \sigma_\theta^2)$ from which actions are sampled during training, while the critic outputs a scalar estimate of the expected return $V^\pi(\mathbf{s})$ from the state, $\mathbf{s}$, under the current policy.

The policy is updated using the PPO clipped objective, which constrains the magnitude of policy updates through a clipping parameter $\epsilon = 0.2$ applied to the probability ratio between successive policies. The advantage function is estimated using Generalized Advantage Estimation (GAE) with $\lambda = 0.95$, and an entropy with coefficient 0.01 is added to encourage exploration during the early stages of training. A detailed description of the PPO algorithm can be found in~\cite{ppo}. The complete set of hyperparameters used in this study is summarized in Table~\ref{tab:hyperparameters_dual}.

In a standard Markov Decision Process (MDP), the current state is assumed to contain all information necessary for optimal decision-making~\cite{sutton1998reinforcement}. In the present experimental system, this assumption is violated by three sources of latency and incomplete information. First, the DC motors exhibit an actuation lag of approximately \SI{200}{ms} corresponding to two sampling intervals at the \SI{10}{Hz} control frequency between command input and full-speed operation. As a consequence, the effect of a control action is not immediately reflected in the observed displacement, and the agent must associate past commands with their delayed kinematic outcomes. Second, the displacement sensors provide only instantaneous position measurements. Velocity and acceleration, which are essential for characterizing the oscillatory dynamics, must be inferred from the temporal evolution of the position signal. Third, and unique to the tandem body configuration, the upstream actuation influences the downstream cylinder only after the modified wake convects across the cylinder gap. At the present spacing of, $L$, and free-stream velocity, $U$, this wake convection introduces an additional time delay on the order of $\tau = L/U$, during which the downstream displacement carries no information about the recent upstream control action.

Together, these three sources of latency render the problem a Partially Observable Markov Decision Process (POMDP)~\cite{singh1994learning}. A well-established approach to mitigating partial observability is to augment the state vector with a finite history of past observations and actions, forming a sufficient statistic that approximates the belief state~\cite{bertsekas2012dynamic}. This strategy has demonstrated success in robotics~\cite{dulac2021challenges} and has recently been adopted in DRL-based active flow control~\cite{xia2024active, weissenbacher2025reinforcement}. In the present work, we retain the most recent $T$ time steps of displacement and action data for each cylinder, yielding the state vector:
\begin{equation}
\mathbf{s} = \left[\mathbf{y}_u^{\text{hist}},\; 
\mathbf{y}_d^{\text{hist}},\; \mathbf{A}_u^{\text{hist}},\; 
\mathbf{A}_d^{\text{hist}}\right],
\end{equation}
where
\begin{equation}
\mathbf{y}_i^{\text{hist}} = \left[\frac{Y_{i,t-(T-1)}}{D},\, 
\dots,\, \frac{Y_{i,t}}{D}\right], \qquad
\mathbf{A}_i^{\text{hist}} = \left[a_{i,t-(T-1)},\, \dots,\, 
a_{i,t}\right],
\end{equation}
represent the displacement history (normalized by the cylinder diameter $D$) and action history of cylinder $i$, respectively, resulting in a state vector of dimension $4T$. The history window $T$ is treated as a design parameter. At a control frequency of \SI{10}{Hz}, a window of $T = 15$ steps spans approximately $\tau_T = \SI{1.5}{s}$ of physical time. Here, $\tau_T$ is set to $2\tau$, such that it is sufficient for the policy network to infer the delayed motor responses with their commands, and to capture the wake convection delay between upstream actuation and its downstream effect.

At each time step, the agent outputs a two-dimensional action vector $\mathbf{a} = [a_{u},\, a_{d}] \in [-1, 1]$, which is mapped to the angular velocities $\Omega_u$ and $\Omega_d$ of the upstream and downstream motors, respectively. In the underactuated case, when only the upstream cylinder is actuated, the policy outputs a single action, $a_u$, and the downstream command is fixed at $a_d = 0$.

The reward at each time step is defined as the negative weighted sum of the root-mean-square displacements computed over a sliding window of $N_w$ time steps:
\begin{equation}
r = -w_u \, \mathrm{RMS}_{N_w}(y_u) 
      - w_d \, \mathrm{RMS}_{N_w}(y_d),
\label{eq:reward_rms}
\end{equation}
where $w_u$ and $w_d$ are weighting coefficients that govern the relative importance assigned to the suppression of each cylinder. The RMS formulation provides a smoothed measure of vibration amplitude that is less sensitive to instantaneous fluctuations. The influence of the window size $N_w$ on learning performance is examined in the hyperparameter sensitivity study presented in Section~\ref{sec:full}. Also, as will be demonstrated, 
the choice of $w_u$ and $w_d$ has a profound effect on the emergent control strategy, particularly in the underactuated configuration where the two system is not fully controllable.

Training is conducted directly on the physical experimental system in real time. The agent interacts with the hardware in episodes of $N = 256$ time steps at a control frequency of \SI{10}{Hz}, yielding an episode duration of approximately \SI{25.6}{s}. At each time step, the agent observes the current 
state, $\mathbf{s}$, samples an action from the stochastic policy $\pi_\theta(\mathbf{a}|\mathbf{s})$, transmits the corresponding motor commands, $\mathbf{a_i}$, and receives the scalar reward, $r$. At the end of each episode, the policy and value networks are updated using the collected trajectory. Unlike simulated environments, the physical system cannot be reset to arbitrary initial conditions between episodes. Each episode begins from the terminal state of the previous one.

After training converges, the stochastic exploration policy is replaced by its deterministic mean:
\begin{equation}
\mathbf{a}_i = \mu_\theta(\mathbf{s}),
\end{equation}
and the network weights are frozen. The deterministic policy is evaluated over an extended testing period of 1{,}000 time steps (approximately \SI{100}{s}) to assess steady-state performance. The effectiveness of the learned control strategy is quantified by the displacement reduction for each cylinder:
\begin{equation}
\Delta_i = \left(1 - 
\frac{\mathrm{RMS}(Y_i)_{\text{controlled}}}
{\mathrm{RMS}(Y_i)_{\text{baseline}}}\right) \times 100\%,
\end{equation}
where the baseline corresponds to the uncontrolled vibration amplitude.

The PPO algorithm is implemented using Stable-Baselines3 (version 2.6.0) built on PyTorch (version 2.7.0). The training environment is custom-developed in Python (version 3.10.0) to interface directly with the experimental hardware. 

\begin{table}[t]
\centering
\caption{Hyperparameters used in the PPO implementation for 
the tandem cylinder control.}
\label{tab:hyperparameters_dual}
\begin{tabular}{ll}
\toprule
\textbf{Parameter} & \textbf{Value / Setting} \\
\midrule
Policy network architecture & $256 \times 256$ (tanh) \\
Value network architecture  & $256 \times 256$ (tanh) \\
Steps per episode ($N$) & 256 \\
Episode duration & $\approx 25.6$~s \\
Batch size & 128 \\
Number of epochs & 10 \\
Discount factor ($\gamma$) & 0.99 \\
GAE parameter ($\lambda$) & 0.95 \\
PPO clipping factor ($\epsilon$) & 0.2 \\
Optimizer & Adam \\
Learning rate ($l_r$) & $2\times10^{-4}$ \\
Weight initialization & Orthogonal \\
Entropy coefficient & 0.01 \\
Action bounds (normalized) & $[-1,\, 1]^2$ \\
Control frequency & 10~Hz \\
\bottomrule
\end{tabular}
\end{table}

\section{Results and Discussion}
\label{sec:results}
The results are organized in order of decreasing actuation authority. We begin with the fully actuated configuration, which establishes a performance baseline, before removing the downstream actuator to investigate underactuated control in both inline and staggered arrangements. In each case, we examine the emergent control strategy discovered by the DRL agent and its physical interpretation.

\subsection{Full actuation in inline configuration}
\label{sec:full}
This section presents the results of the full actuation configuration, in which two motors are simultaneously controlled by a single DRL agent. The agent receives displacement measurements from the vibrating cylinders and outputs the PWM signals to both motors at each time step. The reward function weights both cylinders equally, with $w_u = w_d = 0.5$, such that the agent is incentivized to suppress vibration in both bodies equally. The analysis proceeds in three stages. First, a hyperparameter sensitivity study identifies suitable training parameters. Second, the training dynamics are examined to extract physical insight into the emergent control strategy, and third, the converged policy is deployed to evaluate steady-state control performance.

To identify suitable hyperparameters, a sensitivity study was conducted by independently varying the PPO learning rate, $l_r$, and the RMS reward window size, $N_w$. The results are summarised in Table~\ref{tab:hyperparameter_sweep}. The effect of the learning rate was first examined with fixed window size at $N_w = 1$ (Tests 1--3). At the largest learning rate tested ($1 \times 10^{-3}$, Test~1), the agent failed to converge to a meaningful control policy, indicating that excessively large policy updates destabilise the optimisation landscape. Reducing the learning rate to $5 \times 10^{-4}$ (Test~2) allowed convergence, albeit to a suboptimal policy yielding a displacement reduction of 80\%. A further reduction to $2 \times 10^{-4}$ (Test~3) produced an equivalent result, confirming that once the learning rate is sufficiently small to ensure stable convergence, the 80\% reduction represented an attractor under the instantaneous reward feedback.

With the learning rate fixed at the stable value of $2 \times 10^{-4}$, the influence of the reward window size was then examined (Tests 3--7). Increasing the window from $N_w = 1$ to $N_w = 5$ (Test~4) has a pronounced effect, improving the displacement reduction from 80\% to 95\%. This improvement is attributed to the smoothing of the reward signal, which reduces variance in the policy gradient estimates without significantly compromising the relation between actions and their outcomes. Increasing the averaging window further to $N_w = 10$ and $N_w = 20$ (Tests~5 and~6) results in a decline in performance to a suboptimal level of approximately 85\%, suggesting that excessive smoothing weakens the causal link between control actions and the observed rewards. For the largest window considered, $N_w = 30$ (Test~7), the reward signal becomes overly smoothed, preventing the agent from converging successfully.
\begin{table}[tb]
\centering
\caption{Summary of $lr$ and $N_w$ sensitivity experiments. The displacement reduction percentage is computed by comparing the average RMS displacement during the last 20\% of the training session to the uncontrolled experiment. ``NA'' indicates that the agent did not converge to a meaningful policy.}
\label{tab:hyperparameter_sweep}
\begin{tabular}{cccc}
\toprule
Test & $l_r$ & $N_w$ & $\dfrac{\Delta_u + \Delta_d}{2}$ (\%) \\
\midrule
1 & $0.001$ & 1 & NA\\
2 & $0.0005$ & 1  & 85 \\
3 & $0.0002$ & 1  & 85 \\
4 & $0.0002$ & 5  & 95 \\
5 & $0.0002$ & 10 & 85 \\
6 & $0.0002$ & 20 & 85 \\
7 & $0.0002$ & 30 & NA \\
\bottomrule
\end{tabular}
\end{table}

\begin{figure}[tb]
    \centering
    \includegraphics[width=0.80\linewidth]{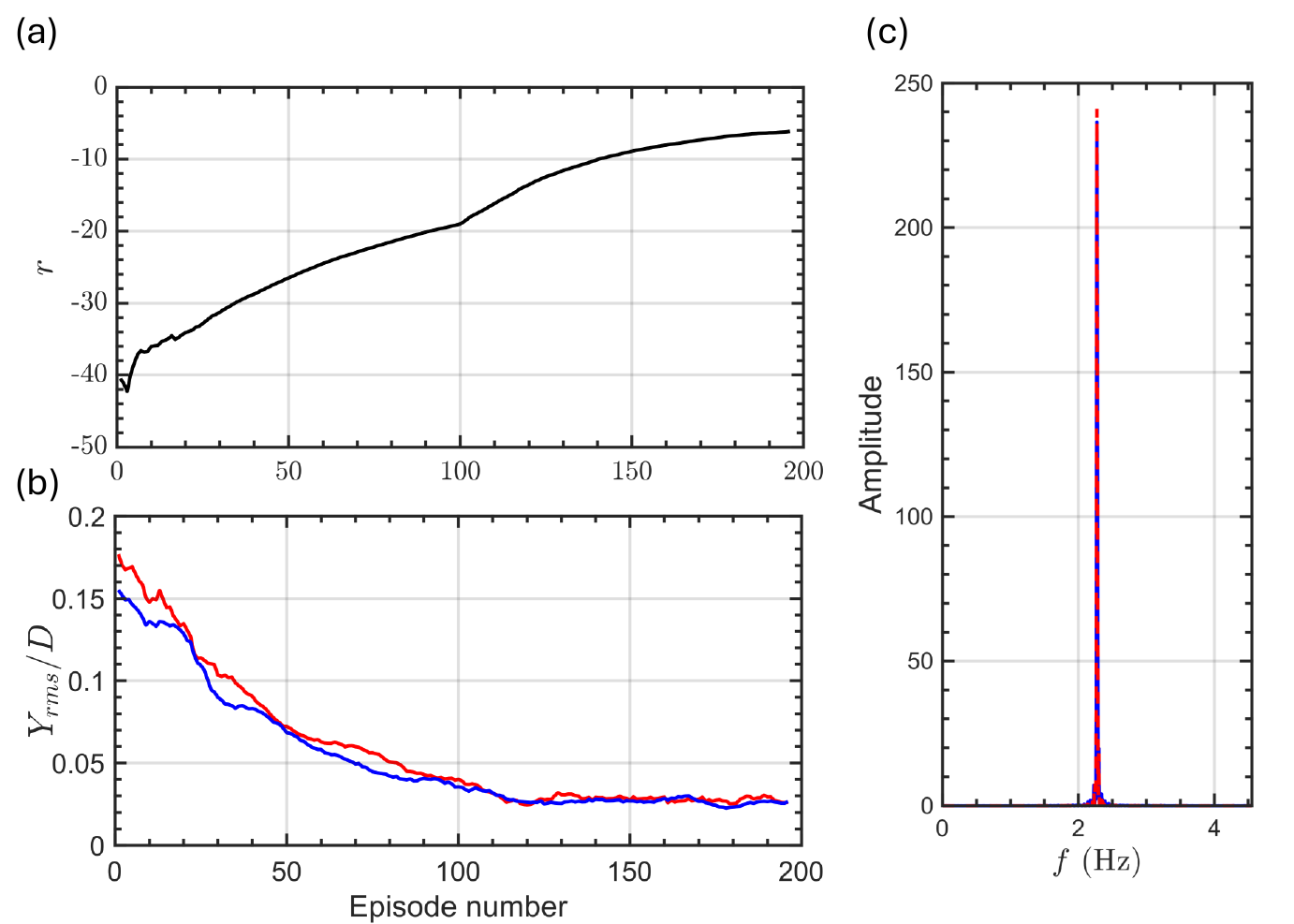}
    \caption{(a) Evolution of the sum of rewards collected over an episode during the training for $l_r = 2 \times 10^{-4}$ and $N_w = 5$. (b) Evolution of the RMS displacement per episode during training. (c) Power spectral density of the motor speed under the converged policy. (red: upstream cylinder, blue: downstream cylinder)}
    \label{fig:direct_reward}
\end{figure}

Having identified the optimal hyperparameter combination ($l_r = 2 \times 10^{-4}$, $N_w = 5$), the training dynamics of Test~4 is inspected in more detail. Figure~\ref{fig:direct_reward}(a) presents the evolution of the training reward over successive episodes. The reward is defined as the cumulative sum of instantaneous rewards over all time steps within an episode. As illustrated in the figure, the reward initially assumes large negative values, reflecting the elevated vibration amplitudes observed during early episodes, when the policy has not yet learned an effective control strategy. As training progresses, the reward increases monotonically before stabilizing at a near constant plateau, indicating that the DRL agent has converged to a stable control policy. Figure~\ref{fig:direct_reward}(b) illustrates the root-mean-square (RMS) displacements, computed within each episode, plotted against the episode number. It can be clearly seen that the amplitude of vibration associated with both cylinders decreases as the policy improves, with both settling at substantially reduced RMS levels when convergence is reached. Finally, Fig.~\ref{fig:direct_reward}(c) illustrates the frequency content of the angular velocities of both motors under the converged policy. The results indicate that the learned control strategy drives both motors to execute steady sinusoidal rotations at a common frequency of approximately $f \approx 2.27~\si{Hz}$, which is markedly higher than the cylinders’ natural frequency, $f_n \approx 1.3$. This elevated operating frequency is a defining feature of the learned control mechanism and will be analyzed in greater detail in the following discussion.

\begin{figure}[tb]
    \centering
    \includegraphics[width=0.80\linewidth]{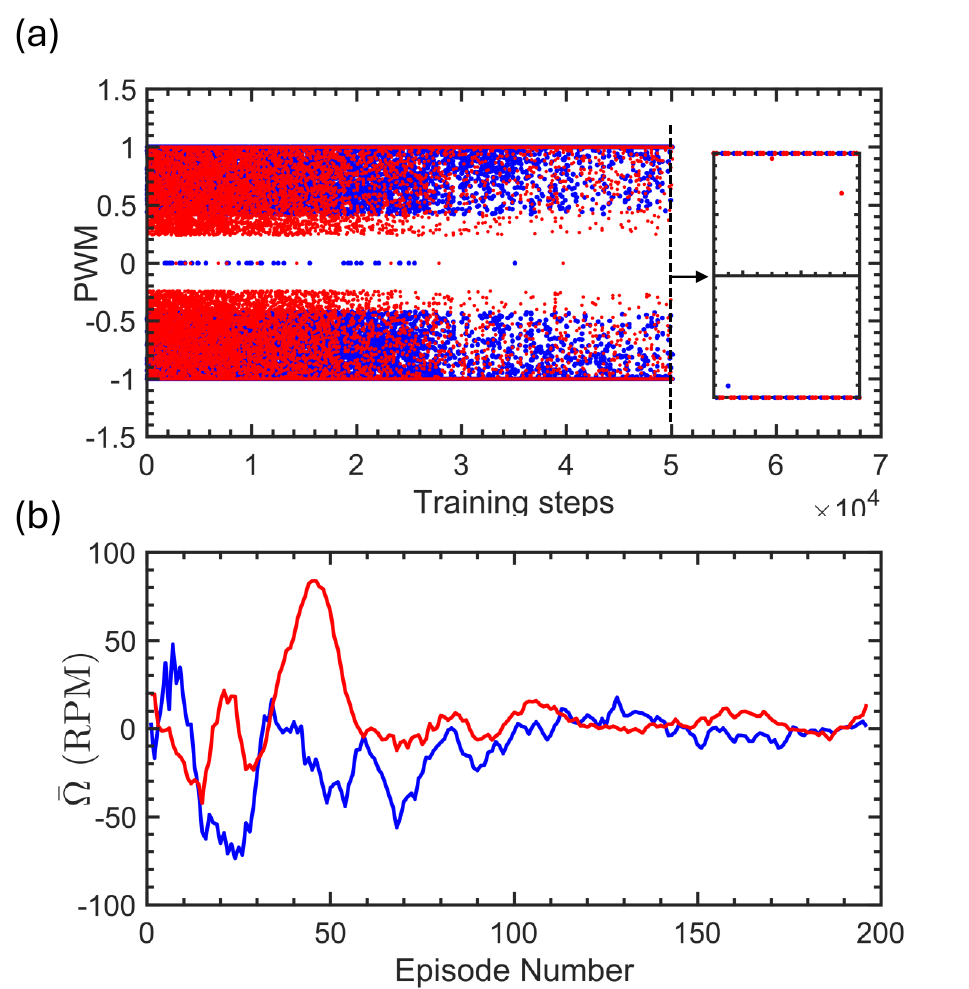}
    \caption{(a) PWM commands issued by the DRL agent to both motors (red: upstream, blue: downstream) over the course of training. (b) Mean rotation speed ($\Bar{\Omega}$) of both motors (red: upstream, blue: downstream) averaged over each training episode.}
    \label{fig:trainActions}
\end{figure}

The nature of the control strategy characterized by the PWM signals of the two motors that emerged during the DRL training is shown in Fig.~\ref{fig:trainActions}(a). During the early episodes, the agent explores the full continuous action space, producing commands distributed broadly across the interval $[-1, 1]$. As training progresses, the action distribution undergoes a transition in which the intermediate values are gradually eliminated, and the commands increasingly concentrate at the saturation limits $\pm 1$. During the final stages of training, the agent converges to a bang-bang control policy~\cite{sonneborn1964bang}, in which both motor commands alternate exclusively between the extreme values $-1$ and $+1$, with very little intermediate commands, as also evident from the inset of Fig.~\ref{fig:trainActions}(a) showing the last 256 interaction steps. 

This convergence to bang-bang actuation is further corroborated by the evolution of the mean of the rotational speed, $\Bar{\Omega}$ shown in Fig.~\ref{fig:trainActions}(b). During the early exploratory phase, the mean values of both motors exhibit large fluctuations, reflecting the agent's broad exploration of the action space with asymmetric and variable commands. As the policy matures, the mean of both motors gradually decays toward zero and stabilizes, confirming that the converged strategy produces a symmetric, zero mean, back and forth motion of both motors. 

\begin{figure}[tb]
    \centering
    \includegraphics[width=0.80\linewidth]{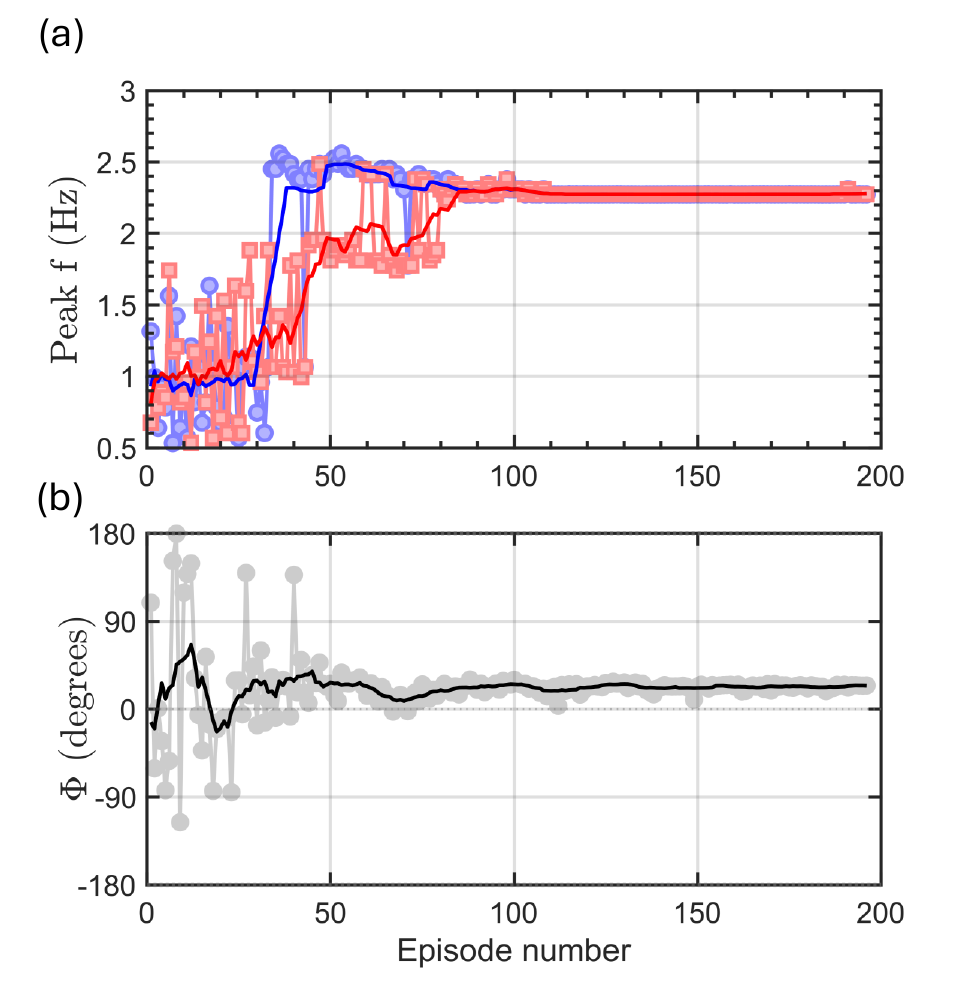}
    \caption{ (a) Evolution of the dominant actuation frequency of both motors (red: upstream, blue: downstream) over the course of training. Thin lines denote raw per-episode values and thick lines represent smoothed data. (b) Phase difference between the control actions in each episode.}
    \label{fig:trainSignal}
\end{figure}

To gain further insight into the interpretability of the emergent control strategy, the evolution of the motors' frequency and the phase difference throughout training are examined in Fig.~\ref{fig:trainSignal}. The dominant frequency is extracted via FFT of the RPM signals, and the inter-motor phase difference is estimated using the Hilbert transform~\cite{feldman2011hilbert}. It can be seen that training can be  broadly decomposed into three distinct regimes. During the first regime (episodes 0--40), both motors operate at irregular frequencies and the phase difference fluctuates widely across the full  range $[-\pi, +\pi]$. This reflects the exploratory behaviour of the agent early in training, where diverse control strategies are sampled without any consistent structure emerging. In the second regime (episodes 40--80), a clear transition is observed. The phase difference narrows substantially, converging toward a restricted range, indicating that the agent has identified a motor synchronisation as a critical factor in reducing structural vibration. Remarkably, this phase locking precedes the stabilization of the motor frequency, suggesting that the agent first learns the correct phase relationship between the two actuators before refining the actuation frequency. Once a consistent phase relationship is established, the agent progressively increases the motor frequency, which is accompanied by a sustained reduction in the displacement of both cylinders, as shown in Fig.~\ref{fig:direct_reward}(b). In the third and final regime (episodes 80 and onwards), both the frequency and phase undergo only minor adjustments, indicating that the policy has largely converged. The fine tuning of the phase difference suggests that the agent continues to exploit small corrections in synchronisation to achieve incremental reductions in vibration amplitude. 

Following training convergence, the policy network weights are frozen and the agent is deployed in deterministic mode, where the stochastic policy is replaced by its mean: $a_i = \mu_\theta(\mathbf{s})$. The deterministic policy is evaluated over an extended testing period of 1000 time steps (approximately 100~s) to assess steady-state performance. During deployment, the system is initialized from its uncontrolled vibrating state, and the agent begins issuing control commands from the first time step. 

\begin{figure}[tb]
    \centering
    \includegraphics[width=0.65\linewidth]{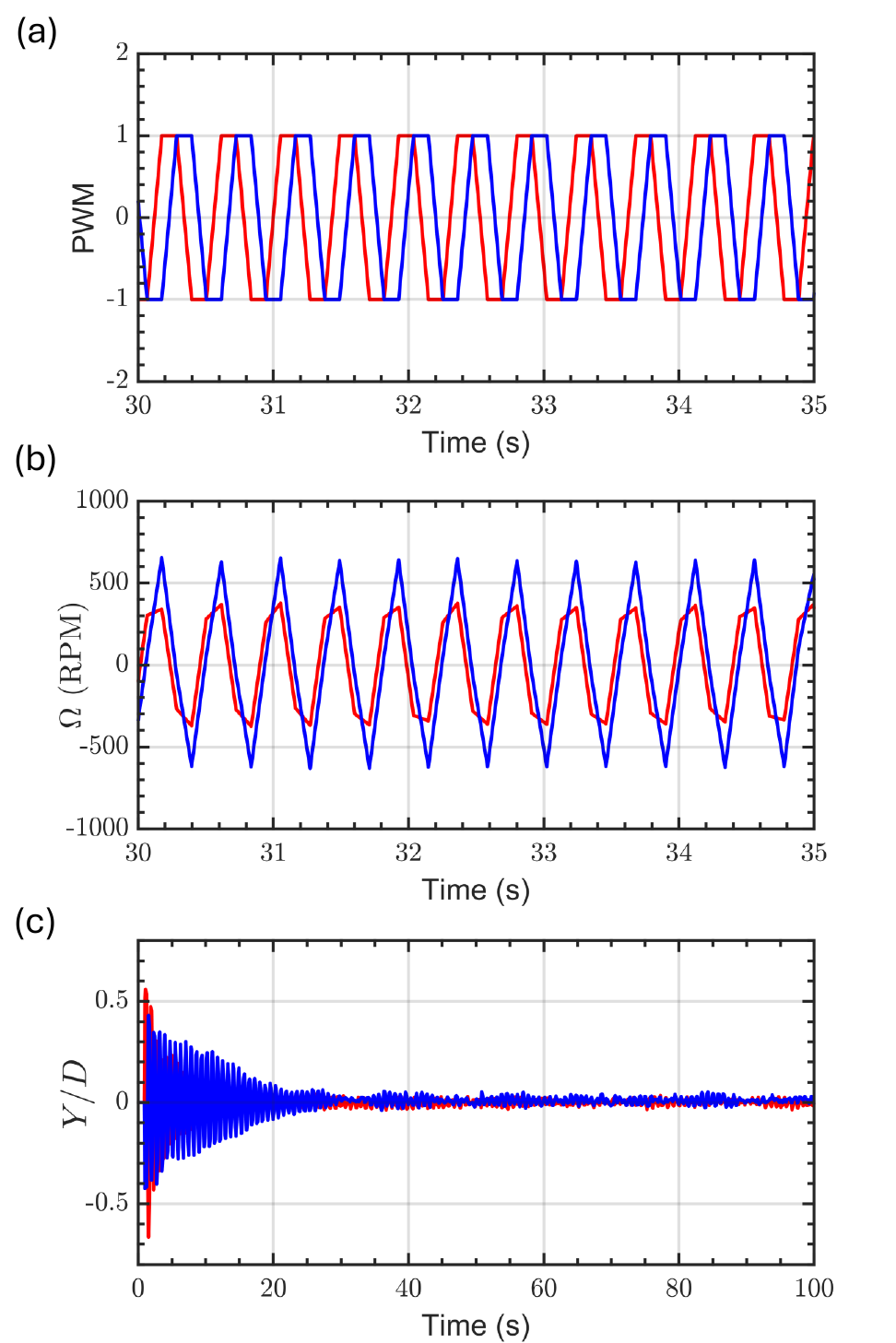}
    \caption{(a) PWM motor signals during deployment (c) Speed of the two motor signals. (c) Dimensionless displacement $Y/D$ of both cylinders during deployment. (Red: upstream, blue: downstream).}
    \label{fig:testing_actions}
\end{figure}

Figure~\ref{fig:testing_actions}(a) presents the PWM signal of both motors commanded by the deployed policy. The agent produces a periodic signal in which both motors alternate between approximately $\pm 1$, confirming the bang-bang nature of the converged strategy observed during training. The two motor commands are phase-shifted relative to one another. The resulting motor rotation speeds are shown in Fig.~\ref{fig:testing_actions}(b). Despite the abrupt switching nature of the voltage commands, the mechanical inertia of the motors produces a sinusoidal waveform in the RPM signal. Both motors rotate at the same frequency $f = 2.27 \si{Hz}$, but with different amplitudes. The upstream cylinder (red) achieves a peak speed of approximately $\pm 600$~rpm while the downstream cylinder (blue) reaches approximately $\pm 400$~rpm. This asymmetry in rotational amplitude arises despite both motors receiving identical PWM commands, and is attributed to differences in the mechanical friction of the two mounting assemblies. Figure~\ref{fig:testing_actions}(c) presents the amplitude of the dimensionless displacement $Y/D$ of both cylinders over the full deployment period. At the onset of control ($t = 0$), both cylinders are vibrating at their uncontrolled amplitudes. Upon activation of the learned policy, both displacement envelopes decay. The amplitude of the upstream cylinder decreases below $Y/D = 0.02$ within approximately 20~s, while the downstream cylinder follows a similar trajectory. By $t \approx 40$~s, both cylinders reach a steady state with attenuated oscillations of $|Y/D| < 0.03$ ($\Delta_u\approx 96.2\% \text{, } \Delta_d\approx 94.4\%$). Importantly, both cylinders are suppressed simultaneously. This confirms that the identified phase-shifted, single-frequency actuation strategy generates a coordinated forcing pattern capable of stabilizing the entire tandem cylinder system.

For the upstream cylinder, the agent appears to have autonomously identified, based solely on reward feedback, that a high-frequency detuning mechanism  suppresses vibrations effectively. This is consistent with prior numerical and experimental findings concerning an isolated cylinder~\cite{du2015suppression, wong2017experimental}. Specifically, the controller enforces a rotation at a frequency of $f \approx 2.27~\si{Hz}$, which is significantly higher than the natural frequency, $f_n \approx 1.3~\si{Hz}$. This imposed rotary motion shifts the vortex shedding frequency to match the forcing frequency rather than the oscillator’s natural frequency, thereby avoiding the resonance responsible for VIV in the largely isolated upstream cylinder.

The downstream cylinder experiences a different FIV mechanism, whereby oscillations emerge from the unsteady interaction between vortices convected from the upstream wake and the shear layers developing on the downstream body. The resulting fluctuating force results in WIV that are sustained only when a favorable coupling between the wake and the shear layer is maintained~\cite{assi2010wake}. It appears that forcing the downstream cylinder to rotate at the same frequency as the upstream cylinder, but with a phase offset, disrupts this coherent coupling between the incoming vortices and the downstream shear layer, thereby suppressing WIV.

\subsection{Underactuated control in an inline configuration}
\label{sec:under}
In the fully actuated configuration, each cylinder is equipped with an independent rotary actuator, affording direct control over both bodies. However, this arrangement may not always be feasible in engineering applications due to energy constraints, cost, or accessibility. We therefore consider the underactuated case, in which only one cylinder is actively controlled while the other is not. Of the two possible configurations, actuating the downstream cylinder is of limited interest, as its rotation primarily modifies the local wake without appreciably altering the flow impinging on the upstream body. Accordingly, in this section we investigate whether a single upstream actuator can simultaneously suppress the vibration of both cylinders, and examine the challenges that arise when training a reinforcement learning policy under this asymmetric control authority. 

\begin{figure}[tb]
    \centering
    \includegraphics[width=0.80\linewidth]{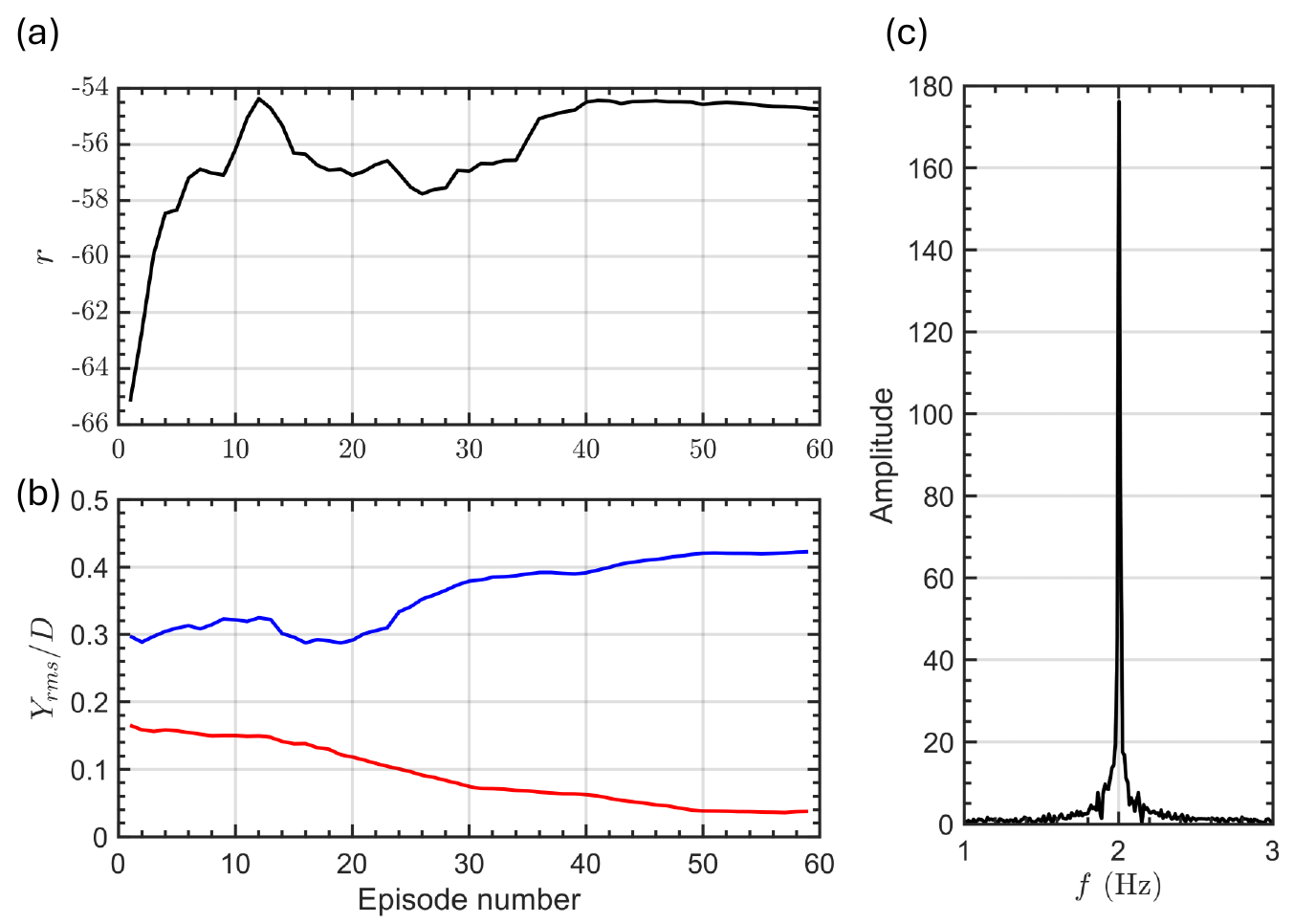}
    \caption{Evolution of the sum of rewards collected over an episode during training of the underactuated inline configuration with equal reward weights ($w_u = w_d = 0.5$).  (b) Evolution of the RMS displacement per episode during training. (red: upstream, blue: downstream). (c) Power spectral density of the upstream motor under the converged policy. }
    \label{fig:underact_equal_reward}
\end{figure}

As an initial attempt, the reward shape from the fully actuated case is retained with equal weighting ($w_u = w_d = 0.5$). As shown in Fig.~\ref{fig:underact_equal_reward}(a), the reward increases rapidly over the first 10 episodes before deteriorating and settling at a low plateau, indicating that the agent discovers a partially effective strategy but fails to sustain improvement. Inspection of
the RMS value of the individual cylinder responses during training (Fig.~\ref{fig:underact_equal_reward}(b)) reveals the cause of this behaviour. The upstream cylinder achieves substantial vibration suppression over the training episodes, whereas the downstream cylinder exhibits the opposite trend, with its amplitude increasing over the course of training. The frequency content of the converged policy, shown in Fig.~\ref{fig:underact_equal_reward}(c), provides further evidence of the partial nature of the learned strategy. The upstream cylinder is rotated sinusoidally at a frequency $f \approx 2.01~\si{Hz}$, well above its natural frequency $f_n$, indicating that the agent has converged on the same frequency-detuning mechanism observed in the fully actuated case. This explains the suppression of the upstream response, but with no actuator available on the downstream body, the coupling that drives the downstream WIV remains unaddressed. 

This asymmetric performance arises from the difference in control authority over the two bodies. The upstream cylinder is directly coupled to the actuator, and consequently the effect of the control action is reflected in the displacement within a single time step. The downstream cylinder, by contrast, is influenced only indirectly through the wake that must convect across the cylinder gap, introducing a substantial delay between action and response. Under equal reward weighting, the agent preferentially exploits the immediate reward than that from the delayed downstream response. This demonstrates that equal reward weighting is insufficient for the underactuated configuration, as it allows the agent to maximize the total reward by suppressing only the directly actuated body.

From a flow perspective, the failure of the DRL agent to converge to an efficient control algorithm in the case of equal weights can be attributed to its reliance on the conventional high-frequency-detuning mechanism of the upstream cylinder to suppress its vibrations. High-frequency vortices arriving from the upstream wake retain sufficient coherence, which according to Assi~et~al.~\cite{assi2010wake} can drive WIV through the same phase-locked vortex--body interaction that operates in the uncontrolled case. Suppressing upstream vibration is therefore not, on its own, sufficient to suppress downstream vibration.

To address the imbalance observed with equal weighting, we revise the reward formulation by assigning asymmetric weights that emphasize downstream suppression, $w_u = 0.2$ and $w_d = 0.8$. The rationale is to penalize downstream vibrations more heavily, thereby forcing the agent to discover control strategies that influence the downstream cylinder through wake modulation rather than focusing on the upstream suppression.

\begin{figure}[tb]
    \centering
    \includegraphics[width=0.80\linewidth]{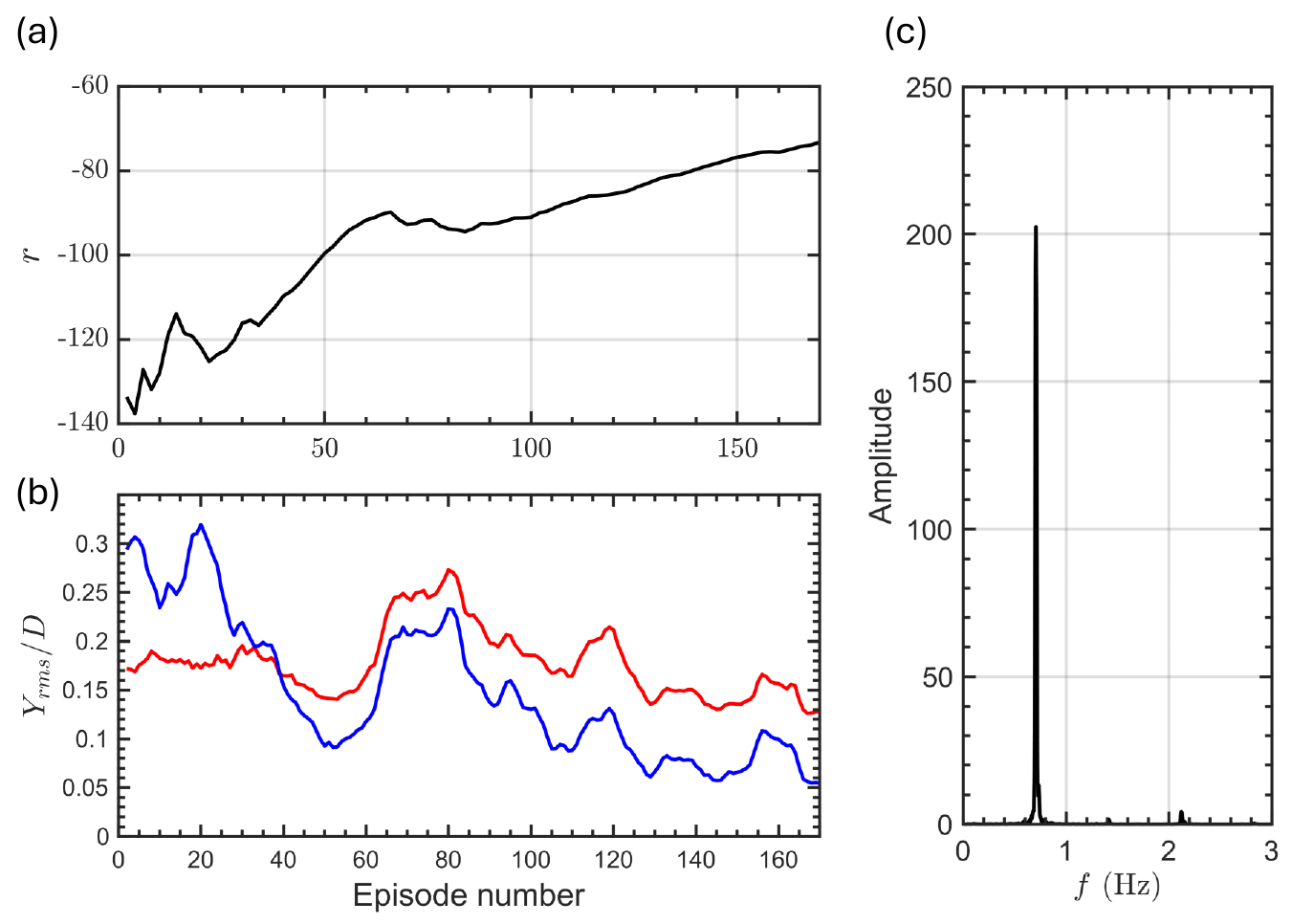}
    \caption{Evolution of the sum of rewards collected over an episode during training of the underactuated inline configuration with equal reward weights ($w_u = 0.2$, $w_d = 0.8$). (b) Evolution of the RMS displacement per episode during training. (red: upstream, blue: downstream). (c) Power spectral density of the upstream motor under the converged policy. }
    \label{fig:underact_asym_reward}
\end{figure}

The training reward under the asymmetric formulation is presented in Fig.~\ref{fig:underact_asym_reward}. The reward exhibits a sustained increase over the full course of training, rising from approximately $r \approx -140$ in the initial episodes to $r \approx -70$ by episode~170. This confirms that the asymmetric weighting successfully redirects the optimization effort toward the downstream cylinder. The corresponding evolution of the RMS 
displacement of each cylinder, shown in Fig.~\ref{fig:underact_asym_reward}(b), confirms that both cylinders 
are now suppressed. For instance, the downstream RMS amplitude settles at approximately $0.05$ and the upstream RMS amplitude settles below $0.15$. The frequency content of the converged policy, shown in Fig.~\ref{fig:underact_asym_reward}(c), reveals a markedly different control strategy from the equal weighting case. The upstream cylinder is now rotated at $f \approx 0.7~\si{Hz}$, well below the natural frequency, $f_n$. 

\begin{figure}[tb]
    \centering
    \includegraphics[width=0.80\linewidth]{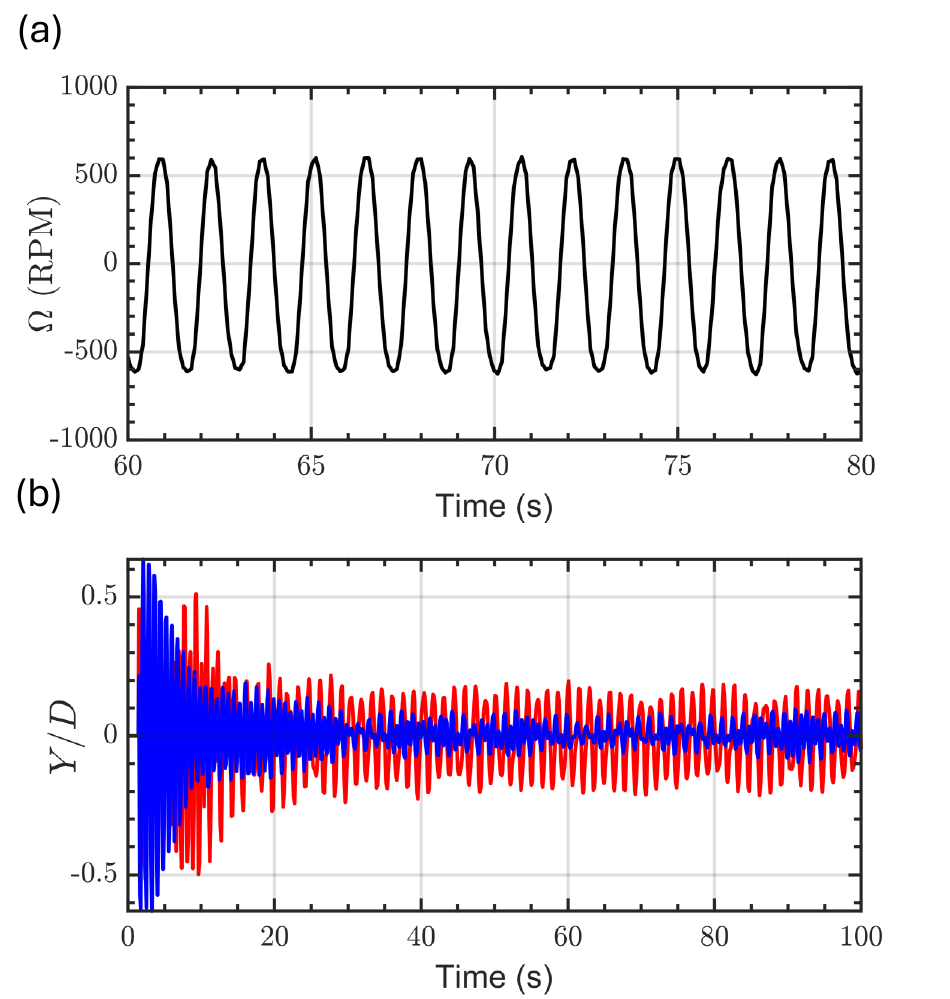}
    \caption{(a) Motor rotation speed during deployment of the underactuated policy.  (b) Dimensionless transverse displacement $Y/D$ of both cylinders during deployment of the underactuated policy with asymmetric weights (red: upstream, blue: downstream).}
    \label{fig:underact_asym_testing}
\end{figure}

Upon deployment of the converged policy, the nature of the discovered control strategy is examined. Figure~\ref{fig:underact_asym_testing}(a) presents the angular velocity of the upstream cylinder. In contrast to the high-frequency pattern observed in the fully actuated case, the rotation speed of the motor oscillates sinusoidally between approximately $\pm 600$~rpm at a low frequency of 0.7~Hz. The displacement time histories of both cylinders under the deployed policy are presented in Fig.~\ref{fig:underact_asym_testing}(b). At the onset of control, the upstream cylinder (red) vibrates at an amplitude of $|Y/D| \approx 0.6$ and the downstream cylinder (blue) at $|Y/D| \approx 0.5$. Both displacement envelopes decay over the first 20--30~s as the control takes effect. The upstream cylinder settles to an amplitude corresponding to a vibration reduction of approximately $\Delta_u = 70\%$, while the downstream cylinder achieves a more pronounced suppression of approximately $\Delta_u = 90\%$. The fact that the downstream cylinder, which receives no direct actuation, is suppressed more effectively than the upstream body is a direct consequence of the asymmetric reward weighting, which biases the optimization toward downstream performance. Nevertheless, the upstream cylinder also benefits substantially from the low-frequency actuation, albeit to a lesser degree. This underscores the ability of DRL to uncover nontrivial, physically meaningful control strategies that would be difficult to uncover using conventional model-based control methods.

The low-frequency rotational oscillation imposed on the upstream cylinder entrains its natural motion and shifts the vortex shedding frequency to a lower value, away from the lock-in region, thereby reducing the amplitude of its vibrations, although less effectively than the high-frequency-detuning mechanism identified in the fully actuated scenario~\cite{du2015suppression, wong2017experimental}. Furthermore, because the modified wake now oscillates at a low frequency of approximately $f \approx 0.7~\si{Hz}$, the vortex arrival rate at the downstream cylinder becomes too slow relative to its intrinsic shear-layer dynamics for the coupling mechanism described by Assi et al.~\cite{assi2010wake} to develop. At this reduced arrival rate, the shear layers associated with the downstream cylinder can complete their roll-up between successive upstream vortex arrivals, thereby preventing the coherent interaction between incoming and locally forming vortices that is necessary to sustain WIV.

\subsection{Underactuated control in a staggered configuration}
\label{sec:offset}
We now consider a more challenging configuration, in which the downstream cylinder is displaced laterally by one diameter ($H/D = 1$) from the centerline of the upstream cylinder, forming a staggered arrangement, see Fig.~\ref{fig:schematic}. This offset introduces additional geometric complexity into the wake interaction, as the downstream body no longer lies directly along the primary wake axis of the upstream cylinder. 

\begin{figure}[tb]
    \centering
    \includegraphics[width=0.80\linewidth]{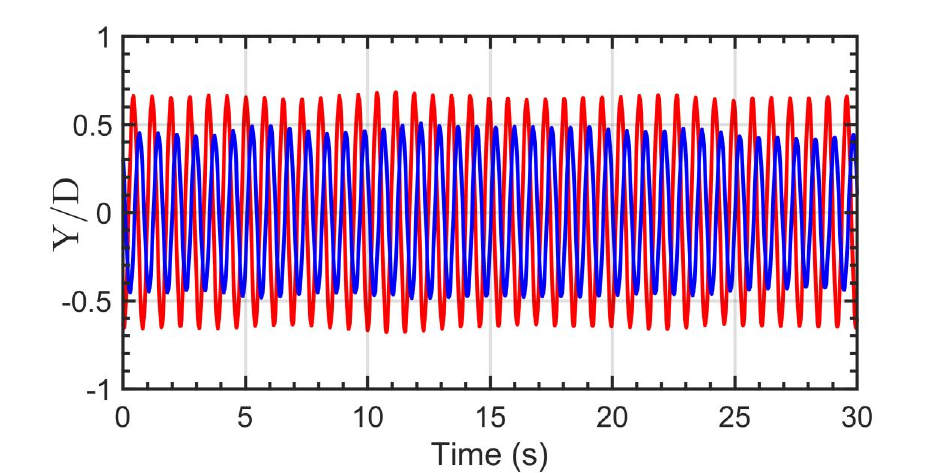}
    \caption{Uncontrolled displacement time histories of the upstream (red) and downstream (blue) cylinders in the staggered configuration at $U = 7$}
    \label{fig:offsetNoControl}
\end{figure}

We first examine the uncontrolled response at the same free stream velocity, $U = 7$ used in the inline configuration. Figure~\ref{fig:offsetNoControl} shows the displacement time histories of both cylinders. The upstream cylinder exhibits essentially identical vibration amplitude to the inline case, as expected given that its wake development is unaffected by the downstream body. However, the downstream cylinder vibrates with a slightly lower amplitude than in the inline arrangement, as the lateral offset weakens the wake-induced forcing.

As an initial attempt to suppress the vibration of both cylinders, the asymmetric reward formulation ($w_u = 0.2$, $w_d = 0.8$) that proved effective in the inline configuration is applied directly here. Under the same DRL hyperparameters, the agent fails to converge to a satisfactory policy. The training reward exhibits persistent oscillations without sustained improvement, and the resulting control strategies do not yield meaningful vibration suppression of either cylinder. This failure is likely attributed to the increased complexity of the offset geometry, in which the lateral displacement weakens the coupling between the upstream actuator and the downstream body. The weaker and more indirect coupling may make reward assignment more difficult for the reinforcement learning agent.

\begin{figure}[tb]
    \centering
    \includegraphics[width=0.85\linewidth]{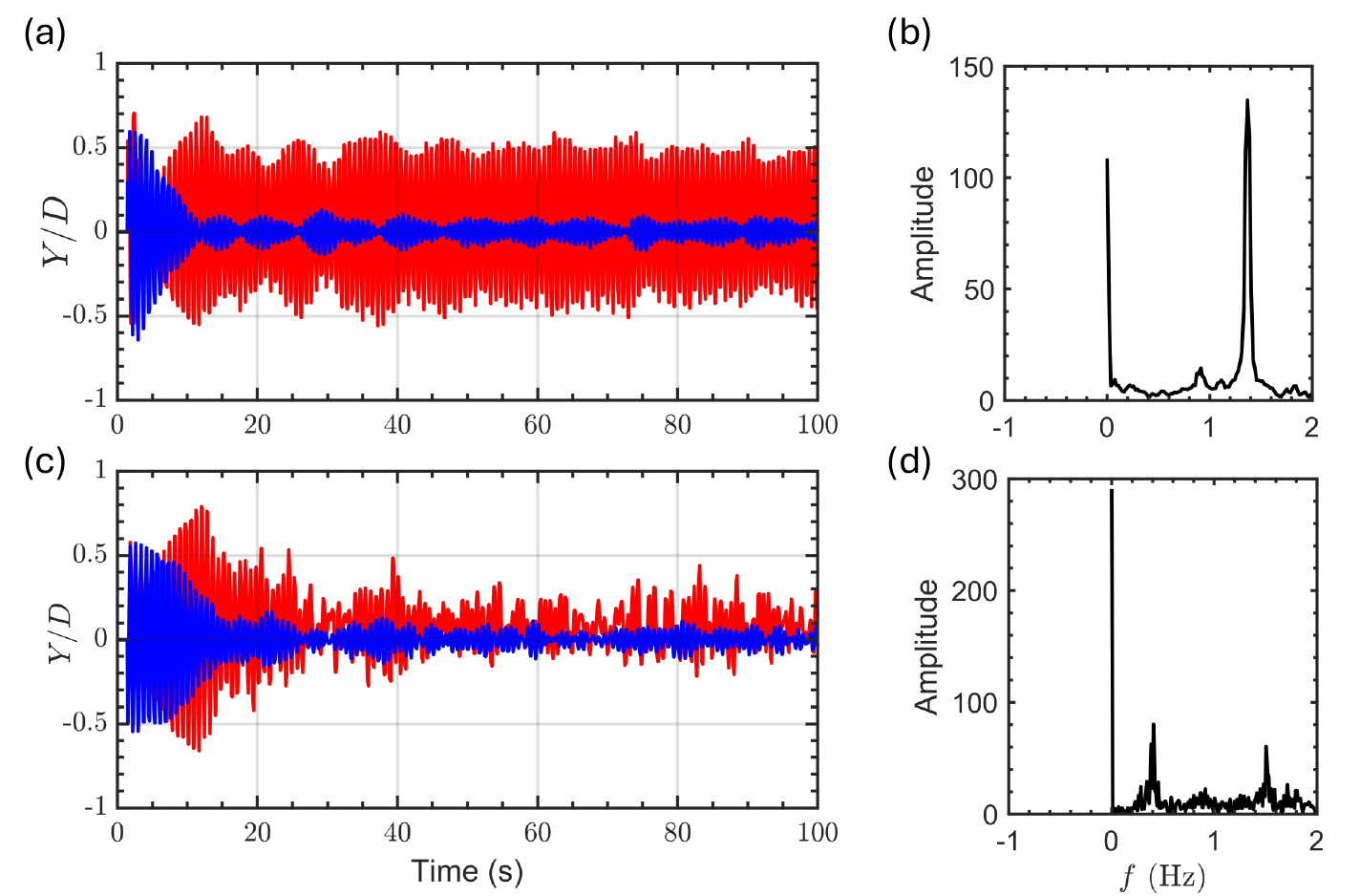}
    \caption{Deployed control performance for the offset underactuated configuration at each curriculum stage.
    (a) stage 1 showing the displacement of the upstream (red) and downstream (blue) cylinders. (b) stage 1 showing the power spectral density of the upstream motor under the converged policy. (c) stage 2 showing the displacement of the upstream (red) and downstream (blue) cylinders. (d) stage 2 showing the power spectral density of the upstream motor under the converged policy. }
    \label{fig:offset_curriculum_time}
\end{figure}

To overcome this difficulty, a curriculum learning strategy~\cite{lin2025survey} is adopted that decomposes the multi-objective control problem into two sequential stages of increasing complexity:
\begin{itemize}
    \item \textbf{Stage 1} ($w_u = 0$, $w_d = 1.0$): The agent is trained with a reward formulation that exclusively penalizes downstream vibrations. By eliminating the upstream suppression objective entirely, the agent can focus on discovering actuation patterns that modulate the upstream wake to stabilize the offset downstream cylinder. 
    
    \item \textbf{Stage 2} ($w_u = 0.2$, $w_d = 0.8$): The upstream displacement penalty is introduced while maintaining the dominant weight on the downstream cylinder. The converged policy from Stage 1 serves as the initialization, so that the agent must now add the upstream suppression onto the established downstream control strategy.
\end{itemize}
For each stage, the agent is trained for 200 episodes, after which the converged policy is deployed in deterministic mode. The resulting time histories under the deployed policy and the associated spectra of the rotating signals are presented in Figs.~\ref{fig:offset_curriculum_time}.

In Stage~1, Fig.~\ref{fig:offset_curriculum_time}(a), the amplitude of the downstream cylinder (blue) decreases to below $0.045D$, corresponding to a displacement reduction of $\Delta_d \approx 85\%$. The upstream cylinder (red), by contrast, maintains high-amplitude oscillations because no penalty is applied to it. 

In Stage~2 (Fig.~\ref{fig:offset_curriculum_time}(b), the introduction of the upstream penalty reduces the upstream cylinder amplitude to approximately $0.12D$ ($\Delta_u \approx 70\%$), while the downstream amplitude is maintained at approximately $0.028D$ ($\Delta_d \approx 89\%$). The agent successfully incorporates upstream suppression without sacrificing the downstream control established in Stage~1. The final converged policy thus achieves simultaneous suppression of both cylinders, despite having direct actuation authority over only the upstream body.

The actuation signal produced by the DRL agent is shown in Fig.~\ref{fig:offset_curriculum_time}(b) for Stage 1 and in Fig. Fig.~\ref{fig:offset_curriculum_time}(d) for Stage 2. In both stages, the resulting motor rotation has two components: an oscillatory component characterized by one or multiple frequencies, and a steady bias at 0~Hz corresponding to a mean rotational speed. 

In Stage~1, Fig.~\ref{fig:offset_curriculum_time}(b), the oscillatory component is dominated by a single frequency near $f \approx 1.35$~Hz, which coincides with the natural frequency of both cylinders, and a steady bias of about $100~\si{rpm}$. The appearance of a steady bias in the control law in the staggered case is very interesting as it is well known in the literature that steady uni-directional rotation of the cylinder helps tilt the mean wake axis towards a non-horizontal line, likely in the direction of the downstream cylinder~\cite{bourguet2014flow}.

In Stage~2, Fig.~\ref{fig:offset_curriculum_time}(d), the actuation signal undergoes a qualitative transformation. The bias increases relative to Stage~1, further enhancing the lateral wake deflection. The oscillatory component, however, now exhibits a composite multi-frequency structure rather than a single dominant peak. Two distinct spectral peaks are identified. A low-frequency component at approximately $0.35$~Hz and a secondary component near $1.5$~Hz.
This dual-frequency structure provides direct evidence that the curriculum learning procedure has produced a control strategy in which distinct frequency components serve distinct physical roles.  Our interpretation is that the low-frequency component addresses the downstream WIV by modifying the coupling mentioned in Section~\ref{sec:under}, while the high-frequency component detunes the upstream VIV through the same high-frequency-mismatch mechanism established for the fully-actuated case. 

\section{Conclusions}
\label{sec:conclusions}
This study presents an experimental implementation of deep reinforcement learning for active suppression of flow-induced vibrations in a tandem cylinder system using rotary actuation. Unlike previous studies, which are numerical in nature and restricted to configurations in which only a single cylinder vibrates or requires control, the present work demonstrates, for the first time, the successful deployment of DRL policies for real-time experimental suppression of simultaneously vibrating tandem cylinders under both fully actuated and underactuated conditions. The key conclusions are summarized as follows.

\begin{itemize}
\item In the fully actuated configuration, a hyperparameter sensitivity study revealed that both the learning rate and reward window size must be carefully selected to achieve convergence. Analysis of the training dynamics 
showed that convergence proceeds along a physically interpretable trajectory. The agent first discovers the correct phase relationship between the two actuators, then progressively refines the actuation frequency, culminating in a high-frequency, phase-locked bang-bang strategy that suppresses the vibrations of both cylinders simultaneously by over 94\%.

\item In the underactuated inline configuration, the reward weighting coefficients proved critical to learning success. Equal weighting causes the agent to exploit the immediate reward from the directly-actuated upstream cylinder while neglecting the indirectly-controlled downstream body. Asymmetric weighting redirects the optimization effort and enables the agent to discover a qualitatively different strategy: a low-frequency rotational oscillation of the upstream cylinder that achieves 70\% and 90\% suppression of the upstream and downstream cylinders, respectively.

\item In the most challenging case concerning underactuated control with a lateral offset between the cylinders, standard training failed to converge, necessitating a curriculum learning approach. A two-stage curriculum produced a control strategy that is unconventional: a statically-biased bi-harmonic sinusoidal motor rotational speed. The presence of a static bias introduces a nonzero mean rotational speed, which likely deflects the upstream wake laterally and may facilitate its interaction with the offset downstream cylinder. Superimposed on this mean motion are two oscillatory components that appear to play complementary roles: a low-frequency component that disrupts resonant coupling with the downstream body, and a higher-frequency component that mitigates the formation of coherent vortices in the upstream wake.
\end{itemize}

Finally, the results of the underactuated configuration carry broader implications. First, the DRL agent autonomously discovers a control strategy that is fundamentally unconventional: a composite bi-harmonic actuation signal with a static bias. This control structure emerges entirely from data-driven interaction, with no prior knowledge of the very complex flow physics. Uncovering such a strategy using classical model-based approaches would require a detailed characterization of the nonlinear wake coupling which maybe unlikely to succeed for complex multi-body configurations. Second, and perhaps more significantly from a practical standpoint, the underactuated framework demonstrates that a single actuator is sufficient to simultaneously suppress the vibrations of two interacting bodies. This finding suggests that in large-scale systems, such as arrays of marine risers, offshore wind turbine clusters, or formations of underwater vehicles, it may not be necessary to actuate every element individually. Instead, by placing actuators on a subset of bodies, one could exploit wake coupling to suppress vibrations across the entire array, substantially reducing the energy consumption and hardware complexity.

\bibliographystyle{elsarticle-num}\biboptions{sort&compress}
\bibliography{mybib}   

%
%
\newpage
\renewcommand{\theequation}{\thesection.\arabic{equation}}
\renewcommand{\thefigure}{\thesection.\arabic{figure}}
\renewcommand{\thesection}{\Alph{section}}
\end{document}
